\documentclass[3p, authoryear]{elsarticle}
\usepackage{amsmath,amssymb,mathrsfs}
\usepackage{stmaryrd}
\usepackage{latexsym}
\usepackage{CJK}
\usepackage{graphicx}
\usepackage{indentfirst}
\usepackage{listings}
\usepackage{xcolor}
\usepackage{booktabs}
\usepackage{stmaryrd}
\usepackage{multirow}
\usepackage{verbatim}
\usepackage{makecell}
\usepackage{lineno,hyperref}
\usepackage{longtable}
\usepackage{hyperref}
\usepackage{upgreek}
\usepackage{mathtools}
\usepackage{float}
\usepackage{caption} 
\hypersetup{colorlinks,breaklinks,
linkcolor=blue,urlcolor=blue,anchorcolor=blue,citecolor=blue}

\newcommand{\uc}{\mathrm{c}}

\newcommand{\txI}{\text{I}}
\newcommand{\rmd}{\mathrm{d}}

\journal{arXiv}

\begin{document}

\title{Application of Rigorous Interface Boundary Conditions \\
in Mesoscale Plasticity Simulations}
\author[cityu]{Jinxin Yu}
\author[hku]{Alfonso H. W. Ngan}
\author[hku]{David J. Srolovitz\corref{cor2}}
\author[cityu]{Jian Han\corref{cor1}}

\address[cityu]{Department of Materials Science and Engineering, City University of Hong Kong, Hong Kong SAR, China}
\address[hku]{Department of Mechanical Engineering, The University of Hong Kong, Pokfulam Road, Hong Kong SAR, China}
\cortext[cor1]{jianhan@cityu.edu.hk}
\cortext[cor2]{srol@hku.hk}
\date{\today}

\begin{abstract}
The interactions between dislocations and interface/grain boundaries, including dislocation absorption,  transmission, and reflection, have garnered significant attention from the research community for their impact on the mechanical properties of materials. 
However, the traditional approaches used to simulate grain boundaries lack physical fidelity and are often incompatible across different simulation methods.
We review a new mesoscale interface boundary condition based on Burgers vector conservation and kinetic dislocation reaction processes. 
The main focus of the paper is to demonstrate how to unify this boundary condition with different  plasticity simulation approaches such as the crystal plasticity finite element, continuum dislocation dynamics, and discrete dislocation dynamics methods.
To validate our interface boundary condition, we implemented simulations using both the crystal plasticity finite element method and a two-dimensional continuum dislocation dynamics model. 
Our results show that our compact and physically realistic interface boundary condition can be easily integrated into multiscale simulation methods and yield novel results consistent with experimental observations.
\end{abstract}
\maketitle

\section{Introduction}
Interfaces play a central role in the plastic deformation of polycrystalline materials. 
Plastic deformation commonly arises from the motion of lattice dislocations, while grain boundaries (GBs) and other interfaces act as barriers to the motion of  dislocations within grains. 
Dislocation pile-ups near  interfaces can induce back stresses within the grains leading to significant  strengthening. 
However, introducing more interfaces into the materials can lead to the classic strength-ductility trade-off; i.e.,  where the materials become stronger but more brittle~\citep{lu2004ultrahigh,zhou2019situ}. 
To overcome this dilemma, materials with nanoscale coherent twin boundaries or heterostructured materials have been the focus of much recent  research~\citep{lu2009strengthening,lu2009revealing,zhu2023heterostructured}. 
The improved performance of these materials may be attributed to tuned dislocation-interface interactions.
For example, coherent twin boundaries can both block dislocations from within the grain as well as serve as dislocation glide planes; stress concentrations near dislocation slip plane-twin boundary intersections can be partially relaxed by such glide.
This demonstrates the importance of a quantitative understanding of the interactions between interfaces and  plasticity in the grain interior as a means of optimizing strength and ductility through interfacial engineering.

 {\it In situ} transmission electron microscopy (TEM) observations show that dislocations from the grain interior may be absorbed into  GBs and transformed (partially) into a set of disconnections on the GB~\citep{kondo2016direct,malyar2017size}. 
These direct observations  provide guidance on how  dislocations from the grain interior interact with interfaces. 
Based on numerous experimental results, the phenomenological Lee-Robertson-Birnbaum (LRB) criteria were proposed~\citep{lee1989prediction,lee1990tem,lee1990situ} to predict the tendency for slip transmission across  GBs. 
However, TEM observations primarily capture individual dislocation-GB interaction events. 
Nonetheless, it remains a challenge  to quantitatively predict the degree of slip transfer, disconnection activity (giving rise to GB sliding), dislocation pile-up, and generation of dislocations on complementary slip systems (giving rise to local hardening) for any particular GB. 

Atomistic simulation (molecular dynamics, MD) can capture dislocation/interface interaction with atomic resolution~\citep{dewald2006multiscale,dewald2007multiscale,zhang2014atomistic}, consisted with  {\it in situ} TEM observations as well as help determine parameters required for larger length and time scale simulation methods~\citep{aragon2022dislocation}.
However,  severe  limitations on the length and time scales accessible to MD make them ineffective for simulating deformation at typical experimentally-relevant  scales. 
This implies the need to perform mesoscale simulations with methods such as discrete dislocation dynamics (DDD)~\citep{lesar2020advances}, continuum dislocation dynamics (CDD)~\citep{el2020continuum} and crystal plasticity finite element method (CPFEM)~\citep{roters2010overview}.

Classically,  most mesoscale simulation methods used to describe plasticity in polycrystalline materials view GBs as blockers of  dislocation motion.
The corresponding GB-interface boundary condition (BC) for dislocation density evolution is a Neumann BC (i.e., no dislocations crossing the GBs); e.g., see~\citep{schulz2014analysis,dunne2007lengthscale,jiang2019effects}. 
However, as discussed above, the interactions between lattice dislocations and interfaces is not this simple. 
Here we examine the application of a more robust interface BC for dislocation/interface interactions  in the context of different mesoscale simulation methodologies.

In continuum dislocation dynamics (CDD), the fundamental degrees of freedom are a set of fields defined throughout the simulation region; the specific choice of field variables differs amongst CDD methods. 
Achaya proposed the field dislocation mechanics (FDM) based on geometrically necessary dislocations (GNDs), as described through the Nye tensor $\boldsymbol{\alpha}$~\citep{acharya2001model,acharya2003driving}.
The dislocation velocity field depends on both the GND  and statistically stored dislocation  (SSD) densities; this is addressed phenomenologically. 
The corresponding grain boundary BCs are formulated by  balancing Burgers vector content in patches of a finitely deforming body~\citep{acharya2007jump}. 
Hochrainer et al.~\citep{hochrainer2007three} proposed a higher-dimensional model which accounts for both the dislocation density and line-curvature fields, later simplifying the approach through phenomenological approximations for the evolution of the average curvature~\citep{hochrainer2014continuum}. 
More recently, another CDD was proposed in which dislocations are represented by vector fields on each slip system~\citep{leung2015new,xia2015computational}. 
The local dislocation line direction is incorporated into the dislocation density and the dislocation density evolution laws at each material point to close the governing equations. 
GBs are considered through  traction and dislocation flux boundary conditions~\citep{el2000boundary,xia2015computational}. 
In this approach, GBs are no longer impenetrable to dislocations and include Burgers vector fluxes  through a source term in the governing dislocation dynamics equations.
Although this type of interface BC  accommodates such issues as Burgers vector conservation, it is not simply  implemented in the corresponding methods and is not compatible with the other CDD methods.  

Crystal plasticity models describe plasticity on a larger scale, providing a means for direct comparisons with macroscopic experimental observations and are commonly implemented within finite element methods, CPFEM~\citep{roters2010overview}. 
Alternative implementations, including the micromechanical self-consistent method~\citep{lebensohn1993self} and the full-field Fourier-based method~\citep{lebensohn2012elasto}. 
At their core, crystal plasticity models, invoke a constitutive law that (in its local formulation) determines the plastic strain rate at each  material point based on the current local material state and the applied load.
CPFEM typically models dislocation-GB interactions through BCs that differ depending on whether the constitutive laws  incorporate strain gradients.
The first type of interface BC introduces an additional slip resistance into the rate equation for plastic slip within the framework of conventional CPFEM~\citep{ma2006consideration,ma2006studying,mayeur2015incorporating}.
Here, special interface elements with larger slip resistance than those within grains are introduced; the proposed additional resistance is based on semi-empirical, rather than physical (dislocation reaction) considerations. 
The second type of interface BC is developed within the framework of strain gradient  theory.
The free energy of slip transfer at the interface is constructed by introducing an additional interface potential to the potential energy functional of the total system~\citep{gudmundson2004unified,aifantis2006interfaces,aifantis2005role}. 
By setting the first variation of the potential energy functional to  zero,  jump conditions for high order tractions across interfaces are obtained; this method can be extended by incorporating more elaborate interface free energy functionals incorporating plastic strain jumps and the average plastic strain across the interface~\citep{fleck2009mathematical}. 
Such BCs do not explicitly  account for crystallographic misorientation across the GBs.
Gurtin proposed a quadratic free energy form for the GB that is essentially an interface defect tensor~\citep{gurtin2008theory}. 
The interface defect tensor measures the GND density at the GBs, analogous to the Nye tensor within the grain interior. 
The flow rules at the interface are derived based on the second law of thermodynamics in the form of a free-energy imbalance. 
This interface BC  couples  grain interiors to  GBs by establishing a microscopic force balance between them. 
Compared with the above methods, this BC can explicitly account for GB crystallography through slip-interaction moduli. 
This BC was numerically implemented within a strain gradient crystal plasticity framework~\citep{ozdemir2014modeling,van2013grain}.
These treatments of GBs can be used for the continuum description of dislocation-GB interactions within a thermodynamically consistent crystal plasticity framework, although the interactions should be viewed as a kinetic rather than a thermodynamic process. 
Note that the second type of BCs is only applicable to strain gradient crystal plasticity and is not compatible with other methods.

The discrete dislocation dynamics (DDD) approach for crystal plasticity tracks the motion of individual dislocation lines within an ensemble of dislocations evolving~\citep{bulatov2006computer}. 
A two-dimensional (2D) DDD model was developed to simulate grain boundary (GB) sliding and the transmission of lattice dislocations across GBs by introducing a Frank-Read source into the grains adjacent to the GB~\citep{quek2014polycrystal,quek2016inverse}. 
While providing valuable insights into the importance of GB sliding during plastic deformation, this model is limited to 2D and relatively small numbers of  discrete dislocations.
Other phenomenological laws based upon the LRB criteria have been proposed to account for individual dislocation-GB interaction in 3D~\citep{cho2020dislocation,zhang2021dislocation}. 
In contrast to the treatment of dislocations as ``points'' in 2D DDD, dislocations are modeled as discrete line segments gliding along  slip planes under a driving force in 3D DDD. 
The kinetic process for such dislocation-GB interactions can be simulated based on prescribed empirical rules.
For example, the process of dislocation transmission through GBs was achieved through a Frank-Read source dislocation bowing out model in GB adjacent grains when the resolved shear stress  on the incoming dislocation exceeded a GB transmission strength~\citep{zhou2012dislocation}. 
Another dislocation-GB interaction model assumed that dislocations can be absorbed into GBs once the resolved shear stress reached a critical value~\citep{zhang2021dislocation} and dislocations can be emitted from GBs when specific rules (dependent on the residual Burgers vector and the resolved shear stress) are satisfied.
Although these dislocation-GB interaction models are intuitive  and easily  implemented within DDD schematics, the empirical criteria are not rigorous and not compatible with CDD methods or the strain gradient plasticity finite element method.
The dislocation-interface BCs discussed above tend to be limited to specific simulation methods. 
In addition, most of these BCs rely on empirical criteria and are not based on rigorous microscopic mechanisms that quantitatively describes reactions between lattice dislocations and interfaces.
Our proposed mesoscale interface boundary condition (BC) is more general and is based upon the kinetics of dislocation reactions at  interfaces~\citep{yu2023mesoscale}. 
Its compact form allows for easy combination with multiscale simulation methods. 
The main challenge is how to relate the input (dislocation density) and output (dislocation flux) quantities to the field variables of the simulation.
In this paper, we first review the mesoscale interface BC in Section \ref{review_BC} and demonstrate the basic idea of its application. 
We then discuss how to implement this interface BC with different CDD simulation methods in Section \ref{CDD} and CPFEM in Section \ref{CPFEM}, respectively. 
For each simulation method, we provide numerical examples that demonstrate the validity of our interface BC.
Finally, we discuss how our interface BC can be naturally extended to  DDD  in Section \ref{DDD}.

\section{The Mesoscale Interface Boundary Condition}\label{review_BC}
\begin{figure*}[bt!]
\centering
\includegraphics[width=0.45\linewidth]{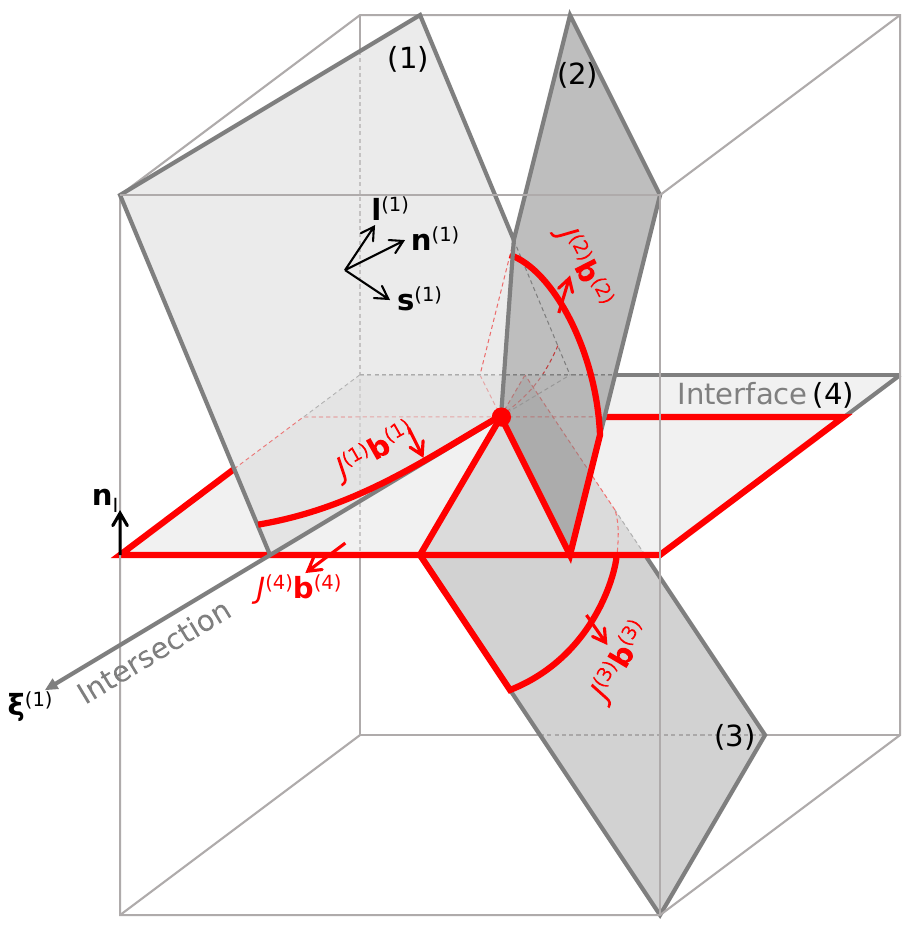}\hspace{-1.78em}%
\caption{\label{fig:bc_3d}
Schematic of single reaction process between the interface and lattice dislocations.
}
\end{figure*}
When  dislocations  from the grain interior approach a GB, they  tend to be blocked, pile up, and or absorbed into the GB.
In order to release the accumulated energy of the GBs,  dislocations may be emitted from the GBs into the same  (reflection) or adjacent  (transmission) grain~\citep{kacher2012quasi,kacher2014dislocation,kacher2014situ}. 
These phenomena are related through dislocation reactions at the interface as shown in the schematic Fig.~\ref{fig:bc_3d}.
We explicitly consider an incoming dislocation slip system in one grain, outgoing slip systems in the original and opposite grains, and slip along the interface.
We  briefly recall the formulation of the mesoscale interface BC~\citep{yu2023mesoscale}.

We first label the slip system $k$ by  superscripts ``$(k)$'',  such that the slip plane unit normal  and slip direction are denoted  $\mathbf{n}^{(k)}$ and $\mathbf{s}^{(k)}$, respectively ($\mathbf{l}^{(k)} \equiv \mathbf{n}^{(k)} \times \mathbf{s}^{(k)}$ is defined to form a local Cartesian coordinate system for each slip system).
When the dislocations approach an interface, the dislocation line $\boldsymbol{\xi}^{(k)}$ will align with intersection slip plane-interface intersection, $\boldsymbol{\xi}^{(k)}_\txI \equiv \mathbf{n}^{(k)}
\times \mathbf{n}_{\txI}$ (see Fig.~\ref{fig:bc_3d}).
In a continuum, the dislocation density $\rho^{(k)}$ at the interface is defined as the number of dislocation lines threading a unit area with unit normal $\boldsymbol{\xi}^{(k)}_\txI$. 
The dislocation flux $J^{(k)}$ is the number of dislocation lines that cross a unit area with unit normal $\boldsymbol{\xi}^{(k)}_\txI$ per unit time. 

To  conserve  Burgers vector, we balance all four Burgers vector fluxes $J^{(k)} (k=1,2,3,4)$ through a   single reaction process (all fluxes are considered positive away form the slip plane-interface intersection should be concealed out at the interface):
\begin{equation}\label{Burgersreaction}
\sum_{k=1}^4 J^{(k)}\mathbf{b}^{(k)} = \mathbf{0}.
\end{equation}
Applying linear response theory, we write the dislocation fluxes as linearly proportional to the driving force (i.e.,  the Peach-Koehler force exerted) acting on the dislocation density.
Combining this with  the conservation of Burgers vector condition Eq.~\eqref{Burgersreaction} allows us to write the interface BC as
\begin{equation}\label{dissipation4}
\mathbf{J}_\txI =
\kappa \mathbf{B}^{-1} (\mathbf{c}\otimes\mathbf{c}) \mathbf{B} \boldsymbol{\tau}_\txI 
\boldsymbol{\rho}_\txI,
\end{equation}
where the subscript ``I'' represents  quantities evaluated at a point on the interface, the Burgers vectors are  $\mathbf{B} \equiv \mathrm{diag}\left(b^{(1)}, b^{(2)}, b^{(3)}, b^{(4)}\right)$, the resolved shear stresses (RSS) are $\boldsymbol{\tau}_\txI \equiv \mathrm{diag}\left(\tau^{(1)}, \tau^{(2)}, \tau^{(3)}, \tau^{(4)}\right)$, and  
$\mathbf{J}_\txI \equiv \left(J^{(1)},J^{(2)},J^{(3)},J^{(4)}\right)^\text{T}$ and $\boldsymbol{\rho}_\txI \equiv \left(\rho^{(1)},\rho^{(2)},\rho^{(3)},\rho^{(4)}\right)^\text{T}$ are the generalized dislocation fluxes and densities. 
The matrix $\mathbf{c} \equiv \left(c^{(234)}, c^{(314)}, c^{(124)}, c^{(132)} \right)^\text{T}$ only depends on the crystallographic orientation of the slip systems meeting at the interface and $c^{(ijk)} \equiv \mathbf{s}^{(i)} \cdot (\mathbf{s}^{(j)} \times \mathbf{s}^{(k)})$ ($i,j,k=1,2,3,4$).
Finally, the scalar quantity $\kappa$ is a reaction constant which is related to the microscopic characteristics of the interface and depends on the detailed atomic structure and bonding at the interface and in dislocation cores.

We now consider the most general case where there are $N$ ($N > 4$) slip systems (see Fig.~\ref{fig:bc_3d}).  
An arbitrary reaction (denoted  ``($n$)'') occurs among any four of the $N$ slip systems.
Similar to Eq.~\eqref{dissipation4}, the kinetics of Reaction ``($n$)'' can be described by
\begin{equation}\label{dissipation_multi}
\mathbf{J}_{(n),\txI}
=
\kappa_{(n)} \mathbf{B}^{-1}_{(n)} (\mathbf{c}\otimes\mathbf{c})_{(n)} \mathbf{B}_{(n)} \boldsymbol{\tau}_{(n),\txI} 
\boldsymbol{\rho}_{(n),\txI},
\end{equation}
$\kappa_{(n)}$ is similar to the $\kappa$ in Eq.~\eqref{dissipation4} and represents the reaction constant for Reaction ``($n$)'' (each reaction will, in general, have a unique reaction constant). 
There are  relations similar to Eq.~\eqref{dissipation_multi} for each reaction.
The total Burgers vector flux, due to all combinations of four reactions $C_{N}^{4}\equiv m$, is
\begin{equation}\label{rbc10}
\bar{\mathbf{J}}_{\txI}
= \sum_{n=1}^{m} \bar{\mathbf{J}}_{(n),\txI}
= \bar{\mathbf{B}}^{-1}
\left(\sum_{n=1}^{m} \kappa_{(n)} \overline{(\mathbf{c}\otimes\mathbf{c})}_{(n)}\right)
\bar{\mathbf{B}}\bar{\mathbf{T}}_{\txI}\bar{\boldsymbol{\rho}}_{\txI}.
\end{equation}
\noindent where 
\begin{align}
 \bar{\mathbf{B}} &\equiv \mathrm{diag}(b^{(1)},b^{(2)}, \cdots,b^{(n)}), \quad
 \bar{\mathbf{T}}_{\txI} \equiv \mathrm{diag}(\tau^{(1)}, \tau^{(2)},\cdots,\tau^{(n)}),  \nonumber  \\
\bar{\boldsymbol{\rho}}_{\txI} &\equiv (\rho^{(1)}, \rho^{(2)},\cdots ,\rho^{(n)})^\text{T},\qquad
\bar{\mathbf{J}}_{\txI} \equiv (J^{(1)}, J^{(2)}, \cdots ,J^{(n)})^\text{T} \nonumber.
\end{align}
This general form of the interface BC is applicable to reactions involving any number of slip systems. 


In our interface boundary condition (BC) Eq.~\eqref{rbc10}, the input and output quantities are the dislocation density $\rho^{(k)}$ and dislocation flux $J^{(k)}$ on each slip system, respectively. The main objective is to establish a connection between these two quantities using field variables appropriate for the different continuum plasticity simulation methods described above.

The first step is to determine the dislocation density at the interface and treat it as an input to our interface BC, as illustrated in Fig.~\eqref{fig:bc_3d}. 
Subsequently, the output dislocation flux $J^{(k)}$ on each slip system is used  to update the plastic strain rate for the CPFEM or the dislocation flux for the CDD method at the interface. 
The proposed interface BC can be employed to determine the outgoing slip system based on  maximum energy dissipation rate in the DDD method. 
Details of the application of the interface BC to various  plasticity methods are discussed below.

\begin{figure*}[t]
\centering
\includegraphics[width=0.55\linewidth]{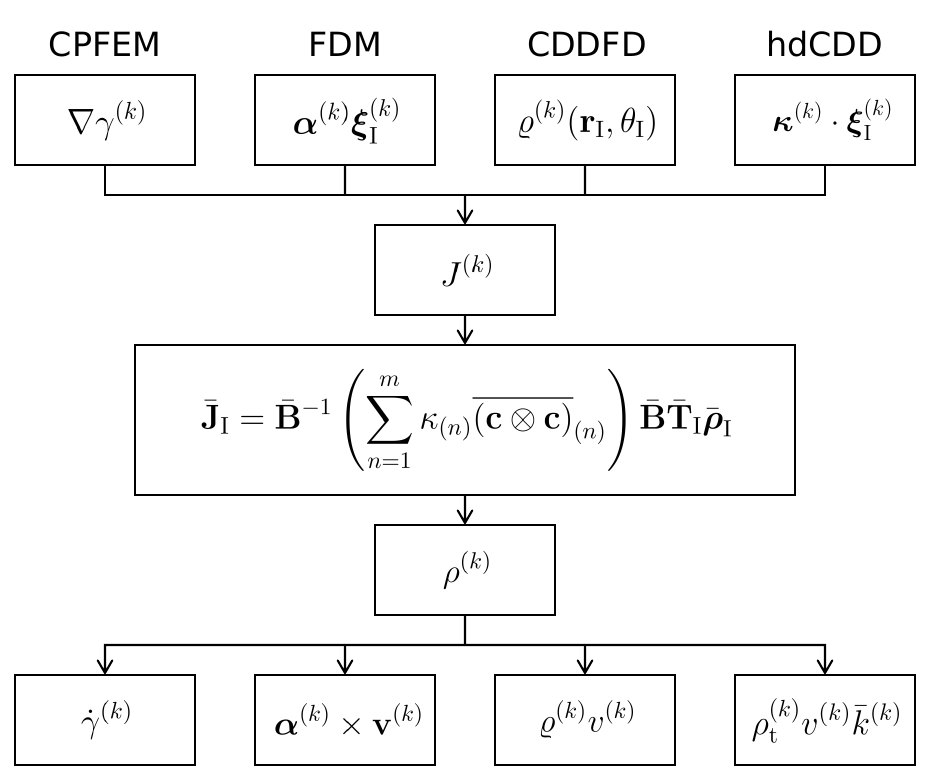}\hspace{-1.78em}%
\caption{\label{fig:cp_flow}
Flow chart showing the procedure for the application of the mesoscale interface BC to different plasticity simulation methods.  
}
\end{figure*}

\section{Application to Continuum Dislocation Dynamics}\label{CDD}
Growing interest in manipulating material properties through microstructure control, grain boundary engineering and interest in nanoscale devices through miniaturization of devices, has increased the community's focus on physically motivated, dislocation-based continuum theories of plasticity~\citep{acharya2001model,leung2016dislocation,hochrainer2007three}. 
In recent years, several  continuum approaches have been introduced that can simulate the motion of curve dislocation lines. 
As described above, several continuum dislocation dynamics (CDD) methods have been developed, based on different choices of field variables, to describe the coarse-grained dislocation density field. 
We consider three such CDD schemes here and show how to combined these with our general interface boundary condition. 

Field dislocation mechanics, a continuum dislocation plasticity approach proposed by Achaya and coworkers~\citep{acharya2001model,acharya2003driving}, focuses on the Nye tensor $\boldsymbol{\alpha}$ (dislocation density tensor) on each slip system as  the primary state variable. 
The Nye tensor is a local measure of the  geometrically necessary dislocation (GND)  density:
\begin{equation}\label{Nye_tensor}
\boldsymbol{\alpha}^{(k)}=\mathbf{b}^{(k)} \otimes \boldsymbol{\rho}^{(k)}, \quad\text{or}\quad {\alpha}^{(k)}_{ij}=b_i^{(k)} \rho_j^{(k)},
\end{equation}
where $\mathbf{b}^{(k)}$ represents the Burgers vector of $k^{\text{th}}$ slip system and $\boldsymbol{\rho}^{(k)}=\rho^{(k)}\boldsymbol{\xi}^{(k)}$ is the density of dislocations   on that slip system with unit tangent vector $\boldsymbol{\xi}^{(k)}$, i.e., $\rho^{(k)}$. 
The Nye tensor for  slip system $k^{\text{th}}$ evolves as 
\begin{equation}\label{Achaya_BC}
\dot{\boldsymbol{\alpha}}^{(k)}=-\nabla \times (\boldsymbol{\alpha}^{(k)} \times \mathbf{v}^{(k)})+\mathbf{g}^{(k)}
=-\nabla \times (\rho^{(k)} v^{(k)} \mathbf{b}^{(k)} \otimes \mathbf{n}^{(k)})+\mathbf{g}^{(k)},
\end{equation}
where $\mathbf{v}^{(k)}$ is the dislocation velocity along its local normal $\boldsymbol{\xi}^{(k)} \times \mathbf{n}^{(k)}$ and $\mathbf{g}^{(k)}$ is a source term on that slip system.
The dislocation density where this slip plane intersects the interface plane is 
$\rho^{(k)}b^{(k)}\mathbf{s}^{(k)}=\boldsymbol{\alpha}^{(k)}\boldsymbol{\xi}^{(k)}_\txI$.
We can substitute this dislocation density $\rho^{(k)}$ at the interface into the interface BC, Eq.~\eqref{rbc10} and 
$\rho^{(k)} v^{(k)}$ in Eq.~\eqref{Achaya_BC} can be updated from the dislocation fluxes $J^{(k)}$ calculated using the interface BC.
Note  that the  updated dislocation flux only applies at the  interface while the dislocation flux within the grain may be calculated using any appropriate bulk kinematic law.

\begin{figure*}[t]
\centering
\includegraphics[width=0.6\linewidth]{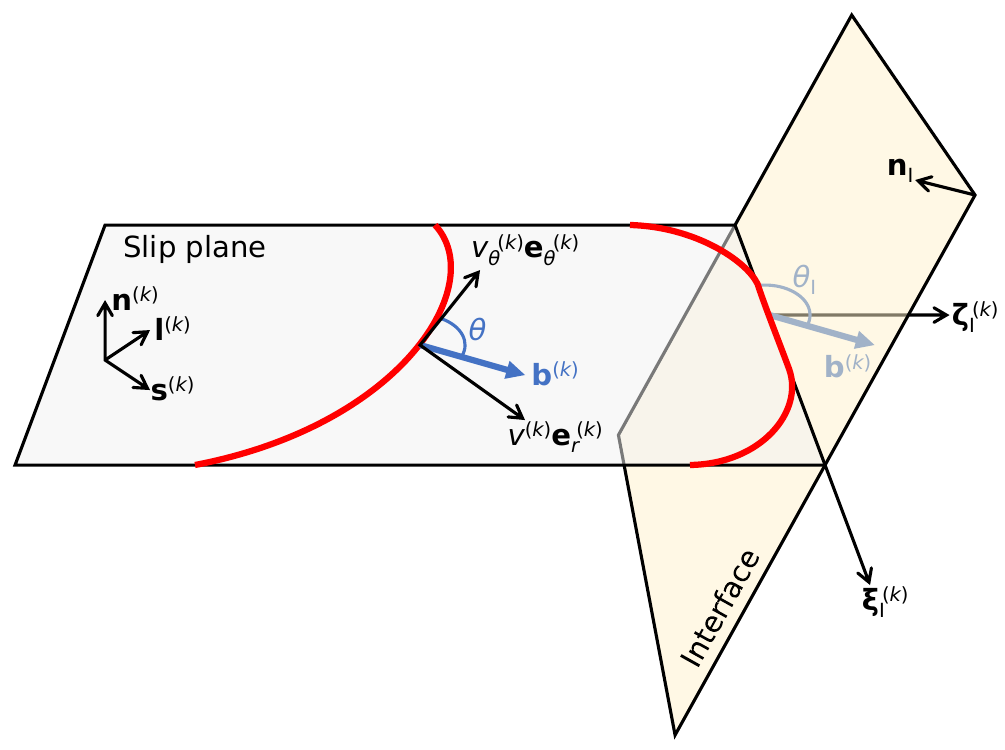}\hspace{-1.78em}%
\caption{\label{cdd_alfanso}
Schematic of dislocation-density function dynamics 
}
\end{figure*}

The continuum dislocation-density function dynamics (CDDFD) approach of Ngan and co-workers~\citep{leung2016dislocation,ngan2017dislocation} considers the dislocation line orientation (relative to the Burgers vector).
Let $\varrho^{(k)}(\mathbf{r},\theta)$ be the density of dislocations with line direction $\mathbf{e}^{(k)}_\theta=\cos\theta \mathbf{s}^{(k)}+\sin\theta \mathbf{l}^{(k)}$ at  $\mathbf{r}$ on slip system $k$.
$\varrho^{(k)}(\mathbf{r},\theta)d\theta$ represents the dislocation density with orientations in the range $\theta$ to $\theta+d\theta$ at  $\mathbf{r}$. 
The dislocation character density   $\varrho^{(k)}(\mathbf{r},\theta)$ evolves as
\begin{equation}
    \dot{\varrho}^{(k)}(\mathbf{r},\theta)=-\nabla(\varrho^{(k)}v^{(k)})\cdot \mathbf{e}^{(k)}_{r}-\frac{\partial (\varrho^{(k)} v_\theta^{(k)})}{\partial \theta},
\end{equation}
where $v^{(k)}$ represents the dislocation  velocity in direction $\mathbf{e}^{(k)}_r \equiv \mathbf{n}^{(k)} \times \mathbf{e}^{(k)}_\theta$ and $v^{(k)}_\theta$ represents the dislocation rotation rate. 
Within the grain, the dislocation velocity  $v^{(k)}$ can be calculated using an appropriate driving-force velocity relation. 
Once  the $v^{(k)}$ field is obtained, the rotational velocity is evaluated as $v^{(k)}_\theta= \rmd v^{(k)}/\rmd e^{(k)}_\theta$, where $e^{(k)}_\theta$ is the dislocation line direction of orientation/character $\theta$.
When the dislocations move to the interface $\mathbf{r}_{\txI}$, the dislocation line will rotate to be collinear with the slip plane/interface intersection $\boldsymbol{\xi}^{(k)}_\txI$ as shown in Fig.~\ref{cdd_alfanso}. 
The dislocation density at the interface on slip system $k$ is $\varrho^{(k)}_{\text{I}}=\varrho^{(k)}(\mathbf{r}_\txI,\theta_\txI)$,
where $\boldsymbol{\xi}^{(k)}_\txI$ has orientation/character $\theta_\txI$. 
Substituting this dislocation density into the interface BC Eq.~\eqref{rbc10}, we  obtain the dislocation flux, $J^{(k)}=\varrho^{(k)} v^{(k)}$.

Hochrainer and coworkers develop a higher-dimensional continuum dislocation dynamics (hdCDD) that includes the dislocation curvature $k(\mathbf{x},\varphi)$ as a second fundamental field variable in addition to dislocation density $\rho(\mathbf{x},\varphi)$~\citep{hochrainer2007three}. 
To illustrate the main idea, consider a two-dimensional case where the dislocations have the same Burgers vector $\mathbf{b}^{(k)}$ moving on $k^{\text{th}}$ slip plane. 
$\rho^{(k)}(\mathbf{x},\varphi)$ represents the dislocation density at  $\mathbf{x}$ on plane $k$,  $\varphi$ is the angle between the dislocation line direction and Burgers vector $\mathbf{b}^{(k)}$ 
(similar to $\theta$ (Fig.~\ref{cdd_alfanso}) in the dislocation-density function dynamics approach~\citep{leung2016dislocation,ngan2017dislocation}).

The original evolution equations for the scalar fields, dislocation density $\rho^{(k)}(\mathbf{x},\varphi)$ and curvature $k^{(k)}(\mathbf{x},\varphi)$, have proven to be computationally costly for simulations. 
This led to the development of simplified approaches based on  Fourier expansions of the local variables $\rho^{(k)}(\mathbf{x},\varphi)$ and curvature $k^{(k)}(\mathbf{x},\varphi)$
~\citep{hochrainer2014continuum,el2020continuum}. 
The zeroth- and first-order coefficients of $\rho^{(k)}(\mathbf{x},\varphi)$ in the Fourier expansion corresponding to the total scalar dislocation density $\rho^{(k)}_{\text{t}}(\mathbf{x})$ and the geometrically necessary density (vector) $\boldsymbol{\kappa}^{(k)}(\mathbf{x})$:
\begin{align}
    \rho^{(k)}_{\text{t}}(\mathbf{x})=\int_{0}^{2\pi}\rho^{(k)}(\mathbf{x},\varphi) \rmd\varphi,\quad 
    \boldsymbol{\kappa}^{(k)}(\mathbf{x})=\int_0^{2\pi}\rho^{(k)}(\mathbf{x},\varphi)
    \left(
    \begin{array}{c}
          \cos\varphi   \\
         \sin\varphi
    \end{array}\right) \rmd\varphi.
\end{align}
The classical Nye tensor is simply recovered as 
    $\boldsymbol{\alpha}^{(k)}=
     \mathbf{b}^{(k)} \otimes  \boldsymbol{\kappa}^{(k)}$.
The evolution of the scalar dislocation density and curvature fields can be expressed as
\begin{equation}\label{hdcdd}
    \left\{
    \begin{array}{l}   
    \dot{\rho}^{(k)}_{\text{t}}=-\nabla \cdot (v^{(k)}\boldsymbol{\kappa}^{(k)}_{\perp})+v^{(k)}\rho^{(k)}_{\text{t}}\bar{k}^{(k)}\\
    \dot{\boldsymbol{\kappa}}^{(k)}=-\nabla \times (\rho^{(k)}_{\text{t}}v^{(k)}\mathbf{n}^{(k)})
    \end{array}\right.,
\end{equation}
where $\mathbf{n}^{(k)}$ is the unit normal of slip plane , $\boldsymbol{\kappa}_{\perp}^{(k)} \equiv \boldsymbol{\kappa}^{(k)} \times \mathbf{n}^{(k)}$ and $\bar{k}^{(k)}$ is the average curvature.
A phenomenological evolution equation for average curvature $\bar{k}^{(k)}$ is required to close the governing equations Eq.~\eqref{hdcdd}. 
Since it is not directly related to our interface BC's application we will not discuss it further.

Similar to the FDM of Achaya in Eq.~\eqref{Achaya_BC}, the dislocation density at the interface is $\boldsymbol{\alpha}\boldsymbol{\xi}^{(k)}_{\txI}=(\mathbf{b}^{(k)}\otimes\boldsymbol{\kappa}^{(k)})\boldsymbol{\xi}^{(k)}_\txI=\rho^{(k)}b^{(k)}\mathbf{s}^{(k)}$, which can be simplified as $\boldsymbol{\kappa}^{(k)}\cdot\boldsymbol{\xi}^{(k)}_\txI=\rho^{(k)}$.
The dislocation flux magnitude is obtained by substituting this into our interface BC Eq.~\eqref{rbc10}.
Again, the  fluxes $\rho_\text{t}^{(k)} v^{(k)}$ and $v^{(k)}\boldsymbol{\kappa}^{(k)}_{\perp}$ in Eq.~\eqref{hdcdd} at the interface can be updated by $J^{(k)}$. 
Since the dislocation lines  align with the interface/slip plane intersection line direction $\boldsymbol{\xi}^{(k)}_{\txI}$ (at the interface), the dislocation curvature  at the interface should be set to $\bar{k}^{(k)}_{\txI}=0$. 

\subsection*{Numerical simulations}
As shown  above, the interface BC Eq.~\eqref{rbc10} can be easily implemented within  different CDD methods; we now present such an application  based upon a relatively simple two-dimensional (2D) CDD model.

\begin{figure}[bt]
\centering
\includegraphics[width=1.0\linewidth]{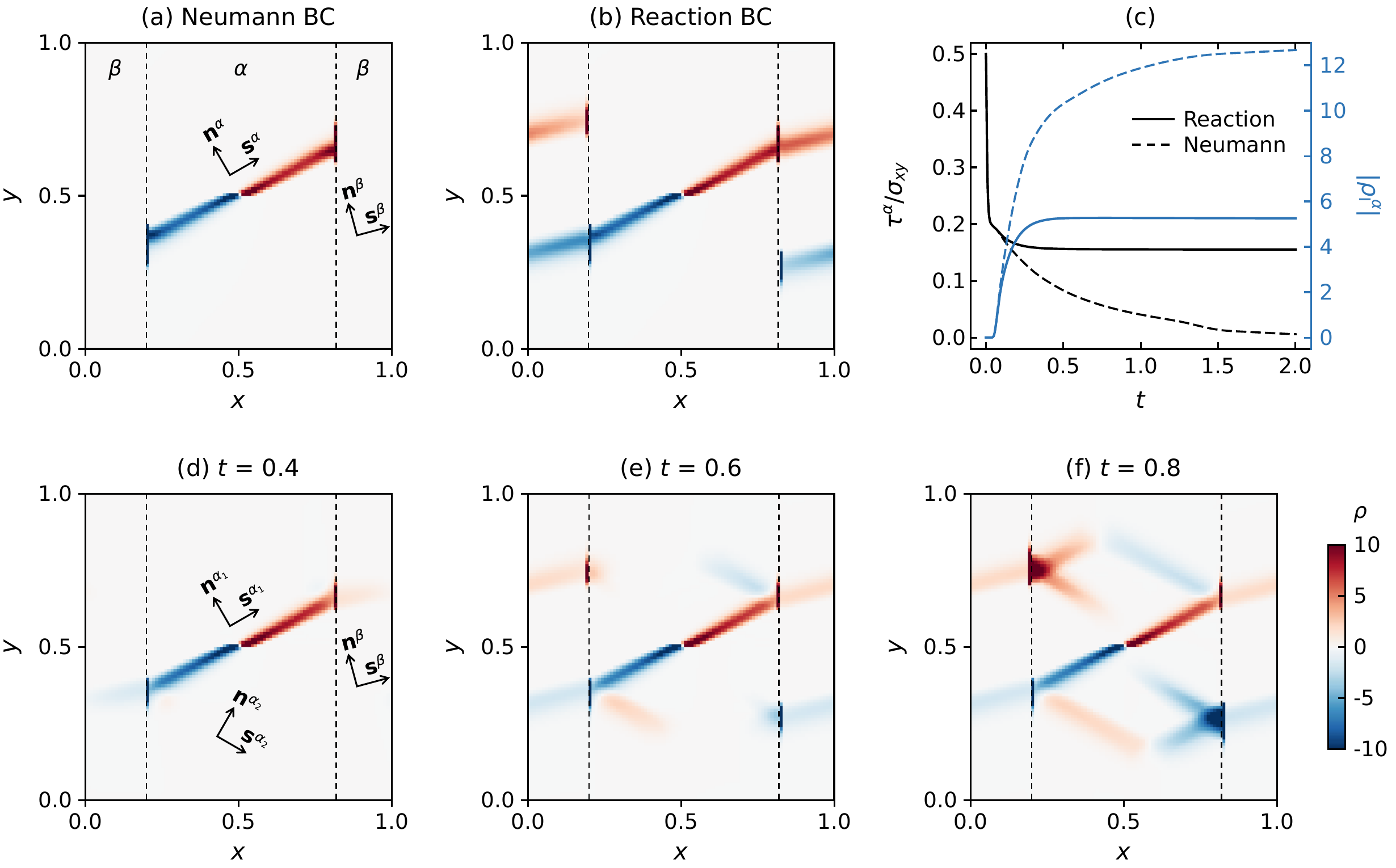}\hspace{-1.78em}%
\caption{\label{fig:cdd_example}
The dislocation density distribution under an applied shear stress $\sigma_{xy}$ with (a) Impenetrable interfaces (i.e., a Neumann BC) and (b) a Reaction BC (i.e., Robin BC)  for a single slip system within each phase (orientations as indicated) where the dashed lines represent the interfaces.
It should be noted that for Impenetrable BC case, it reached equilibrium state while there was no equilibrium or steady state for a Reaction BC case. 
(c) Solid and dashed lines represent the stress at the dislocation source $\tau^{\alpha}$  and the dislocation  density at the interface for the two types of BCs (Impenetrable and Reaction);
(d)-(f) The dislocation distribution under uniaxial loading for the case of a Reaction BC for the case of two slip systems in $\alpha$ and one in $\beta$ at  times, $t=0.4, 0.6, 0.8$.
}
\end{figure}

Consider the lamellar configuration  illustrated in Fig.~\ref{fig:cdd_example}(a) (periodic along the $x$ and $y$-directions). 
There are two phases, $\alpha$ and $\beta$, of width $L_{\alpha}$ and $L_{\beta}$, separated by interfaces (denoted by black dashed lines) in each period. 
The slip direction and  plane normal are $\mathbf{s}^{(k)}$ and $\mathbf{n}^{(k)}$ for $(k) = \alpha$ or $\beta$.
The dislocation density vector is defined as 
$\boldsymbol{\rho}^{(k)} = \rho^{(k)} \boldsymbol{\xi}^{(k)}$ 
($\boldsymbol{\xi}^{(k)} \equiv \mathbf{s}^{(k)} \times \mathbf{n}^{(k)} $ is the dislocation line direction). 
All dislocation lines are straight, perpendicular to the plane,  in the two-dimensional model and its direction is always; i.e, $\boldsymbol{\xi}^{(k)}=\mathbf{e}_x \times \mathbf{e}_y$.
Hence, the three different CDD schemes  become identical and the evolution of dislocation density can be expressed as (Maxwell-Faraday equation):
\begin{align}\label{Faraday}
\dot{\boldsymbol{\rho}}^{(k)}
&= - \nabla \times \left(\boldsymbol{\rho}^{(k)} \times \mathbf{v}^{(k)}\right)=-\left(\cos\theta^{(k)} \frac{\partial(\rho^{(k)} v^{(k)})}{\partial x} + \sin\theta^{(k)} \frac{\partial(\rho^{(k)} v^{(k)})}{\partial y}\right) \mathbf{e}_z, 
\end{align}
where $\mathbf{v}^{(k)}= v^{(k)}\mathbf{s}^{(k)}$ is the dislocation velocity, $\theta^{(k)}$ is the angle between the slip direction $\mathbf{s}^{(k)}$ and $\mathbf{e}_x$. 
Defining the dislocation flux as
$\mathbf{J}^{(k)}=\rho^{(k)}\mathbf{v}^{(k)}$, the evolution of dislocation density simply becomes the dislocation flux divergence:
$\dot{\rho}^{(k)}=-\nabla \cdot \mathbf{J}^{(k)}$.


The magnitude of the  dislocation velocity may be described by a power law:
\begin{equation}\label{kineticslaw}
v^{(k)}_{+/-}= \pm \mathrm{sgn}(\tau^{(k)}) v^{*(k)} \left|\tau^{(k)}/\tau^{*(k)}\right|^n, 
\end{equation}
where dislocation densities of opposite signs are denoted by subscripts ``$\pm$'', $\tau^{*(k)}$  is the slip resistance for $k^{\text{th}}$ slip system, and the constant $n$ depends on the range of stress. 
The dislocation density  satisfies  the balance:
$\dot{\rho}^{(k)}_{+/-}=-\nabla \cdot \mathbf{J}^{(k)}_{+/-}+ \dot{\rho}_{+/-}^{(k),\text{ann}}+ \dot{\rho}_{+/-}^{(k),\text{gen}}$.
As proposed by Arsenlis and co-workers~\citep{arsenlis2004evolution}, dislocation annihilation  occurs when  opposite signed dislocations come within a  capture radius $r_{\uc}$; i.e.,  the annihilation rates are
$\dot{\rho}_{+}^{(k),\text{ann}}
= \dot{\rho}_{-}^{(k),\text{ann}}
= -\rho^{(k)}_{+}\rho^{(k)}_{-}r_{\uc}
|v^{(k)}_{+}-v^{(k)}_{-}|$.
When a stress is applied, a dislocation pair (opposite signs) is emitted from a source at the rate~\citep{kocks1975thermodynamics}
$\dot{\rho}_{+}^{(k),\text{gen}}=\dot{\rho}_{-}^{(k),\text{gen}}=\eta \left|\tau^{(k)}\right|^{m}$,
where $\eta$ and $m$ are constants. 
The net dislocation density is $\rho^{(k)}=\rho^{(k)}_{+}-\rho^{(k)}_{-}$.
The total shear stress  $\tau^{(k)}$ in Eq.~\eqref{kineticslaw} has contributions from both  external  $\boldsymbol{\sigma}^\text{ext}$ and  internal  (associated with all other dislocations) $\boldsymbol{\sigma}^\text{int}$ stresses. 
In the simulations, we employ  reduced variables: 
$\tilde{\rho} \equiv \rho L^2$,
$\tilde{x} \equiv x/L$,
$\tilde{t} \equiv v^* t/ L$, 
$\tilde{v} \equiv v / v^*$, 
$\tilde{\tau} \equiv \tau / K$, 
where $L \equiv L_\alpha+L_\beta$ is the width of the whole cell in $x$ and $K \equiv \mu/[2\pi(1-\nu)]$. 
For simplicity, we  omit the tilde in the reduced quantities  below.
The parameters for these simulations are listed in  Table.~\ref{table:cdd}; since we assume all slip systems have the same properties, the superscript ``$(k)$'' representing  different slip systems are omitted.
\begin{table}[tbp]
\centering
\caption{Parameters employed in the 2D CDD simulations.} 
\renewcommand{\arraystretch}{1.3} 
\begin{tabular}{m{8cm} m{1.5cm}} 
 \hline
 Parameters & Values \\ 
 \hline
 Burgers vector magnitude, $b$ & $1 \times 10^{-3}$ \\ 
 Capture radius for annihilation, $r_{\text{c}}$ & $1 \times 10^{-3}$ \\
 Applied shear stress, $\sigma_{xy}$ & $2 \times 10^{-2}$ \\
 Dislocation source intensity, $\eta$ &  $1.5 \times 10^{7}$ \\
 Dislocation  source exponent, $m$ & 2 \\ 
 Timestep, $\Delta t$  & $5 \times 10^{-4}$ \\ 
 Dislocation  velocity exponent, $n$ & 1 \\
Reference velocity, $v^{*}$ & 1.0 \\
 Slip resistance, $\tau^{*}$ & $1 \times 10^{-2}$\\
 \hline
\end{tabular}
\label{table:cdd}
\end{table}

We first consider a simple case where there is only one slip system within each phase ($\theta^\alpha=30^\circ$ and $\theta^\beta=15^\circ$) under an external shear stress $\sigma_{xy}$ with one operable dislocation source at the centre of phase $\alpha$.
Figure~\ref{fig:cdd_example}(a) shows the equilibrium dislocation density for the case in which the  interface  is impenetrable (i.e., a Neumann BC); the reaction constant $\kappa_{(n)}=0$ in Eq.~\eqref{dissipation_multi}. 
Here, the dislocations glide along the slip plane and pile up near the interface.
When dislocation reactions can occur at the interface ($\kappa_{(n)} \neq 0$), with the interface boundary condition Eq.~\eqref{rbc10} applied, the model approaches a state as shown in Fig.~\ref{fig:cdd_example}(b). 
For this case (constant external stress with a Robin BC), there is no equilibrium or steady state; i.e., the dislocation point source continues to operate such that  dislocations continue to cross the GBs.
In this case, dislocations flow across the interface from the $\alpha$ phase into $\beta$.
Figure~\ref{fig:cdd_example}(c) shows that the pile up dislocation density at the interface is much larger in the case of the impenetrable interface than when the Reaction BC is applied; the Reaction boundary condition makes the interface penetrable. When the interface is impenetrable, the total resolved shear stress (RSS) at the dislocation source in $\alpha$ tends to zero as a result of the back stress from the pile up cancelling the stress at the source -- effectively shutting down the dislocation source.
On the other hand, when the Reaction BC is applied at the interface, the penetrability of the interface makes the pile up weaker and unable to cancel the applied stress -- hence the stress at the source asymptotes to a non-zero value. 
Figure~\ref{fig:cdd_example}(d)-(f) illustrate the evolution of the  dislocation distribution for a case where  there are two slip systems in $\alpha$ phase ($\theta^{\alpha_1/\alpha_2}=\pm30^\circ$) and one in $\beta$  ($\theta^{\beta}=15^\circ$). 
The simulation results demonstrate that the dislocation pile up  from the $\alpha_1$ slip system not only can transmit into $\beta$ phase, but also can be reflected back into $\alpha_2$ slip system. (Note that there is an effect of the periodic BC in the simulation cell.)
These simulation results are in excellent agreement with experimental observations.

\begin{figure}[!t]
\centering
\includegraphics[width=0.75\linewidth]{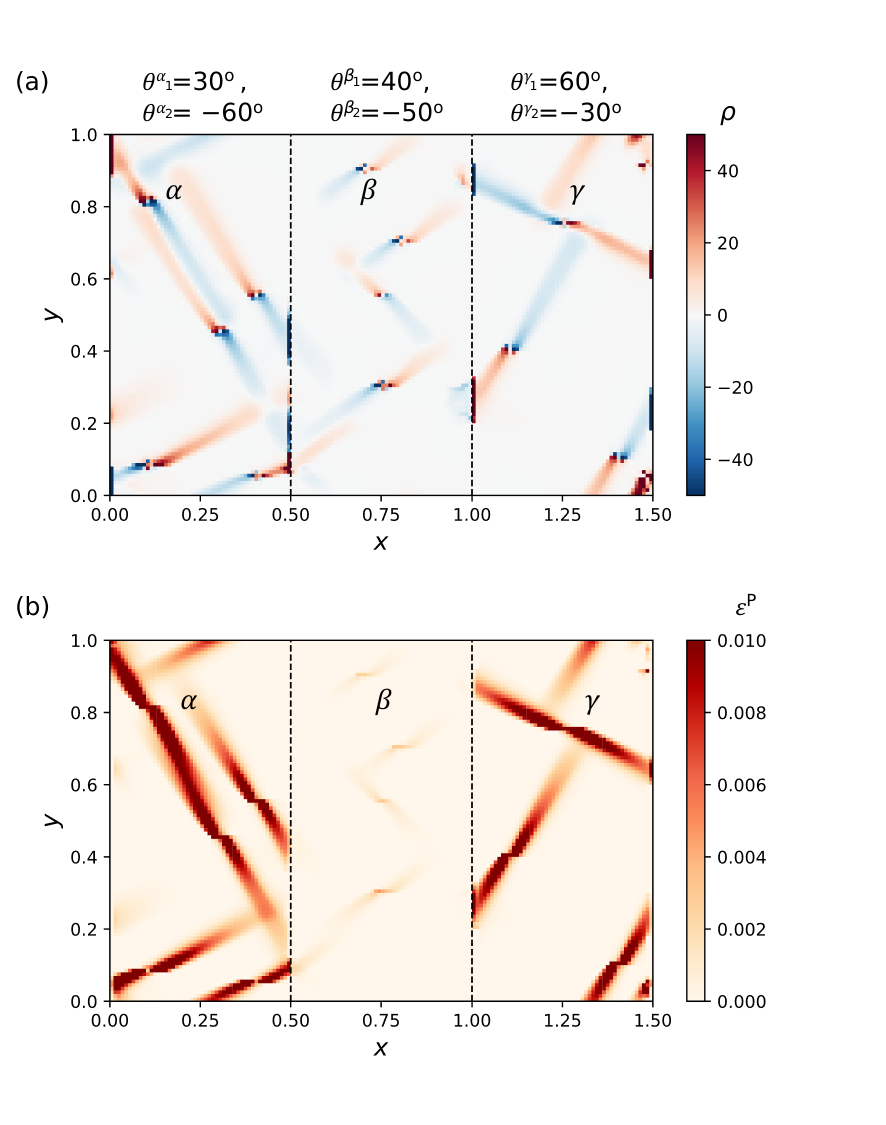}\hspace{-1.78em}%
\caption{\label{fig:cdd_multiple_source}
A three grain lamellar microstructure  with two slip systems and multiple sources within each grain, subject to an external shear stress $\sigma_{xy}$.
(a) The dislocation density distribution. (b) The corresponding von Mises plastic strain $\epsilon^{\text{p}}$.
}
\end{figure}

To illustrate the generality of our interface BC, we  consider a more complex example involving three grains (denoted  $\alpha$, $\beta$, $\gamma$). 
Each grain has two slip systems with the following orientations: $\theta^{\alpha}=30^\circ,-60^\circ$; $\theta^{\beta}=40^\circ,-50^\circ$; $\theta^{\gamma}=60^\circ,-30^\circ$.
In these simulation, the external stress is $\sigma_{xy}$. 
The dislocation sources in $\beta$  are less active since the resolved shear stress on the $\beta$ slip system is small relative to the other  grains. 
For the same reason, it is difficult for slip to propagate from $\alpha$ and $\gamma$ into $\beta$; see Fig.~\ref{fig:cdd_multiple_source}(a).
In this scenario, only a small number of dislocations react at  GBs and most of the slip is confined to the individual grains. 
The primary mechanism responsible for the plastic deformation of this tricrystal is the motion of dislocations within the grains, as depicted in Fig.~\ref{fig:cdd_multiple_source}(b).

\section{Application to the Crystal Plasticity Finite Element Method}\label{CPFEM}

Crystal plasticity methods are robust,  widely-used computational tools for  examining the relationship between mechanical properties and material structure which have proven especially effective in micromechanics applications, including strain hardening in single crystals and texture evolution in polycrystalline aggregates~\citep{roters2010overview}.
Crystal plasticity models are based upon the idea that plastic flow occurs via slip or shear on activated slip systems characterized by a critical resolved shear stress and/or a hardening law. 
Within the framework of small deformation, the plastic distortion is the sum of slip $\gamma^{(k)}$ on  individual slip systems:
\begin{equation}    
\mathbf{H}^{\text{p}}=\sum_{k}\gamma^{{(k)}} \mathbf{s}^{{(k)}} \otimes \mathbf{n}^{(k)}.
\end{equation}
Conventional CPFEM has no description of dislocation physics. 
A Burgers tensor can be defined to link dislocation slip to the plastic strain gradient~\citep{gurtin2010mechanics}:
\begin{equation}\label{burgers_tensor}
    \mathbf{G} = \nabla\times \mathbf{H}^{\text{p}}=\sum_{k}(\nabla \gamma^{{(k)}} \times \mathbf{n}^{{(k)}}) \otimes \mathbf{s}^{{(k)}},
\end{equation}
where $\mathbf{G}^{\text{T}} \mathbf{e}$ represents the Burgers vector (per unit area) for infinitesimal closed circuits on any plane with unit normal $\mathbf{e}$. 
For any slip system ${k}$, $\mathbf{s}^{(k)}$ and
$\mathbf{l}^{(k)} \equiv \mathbf{n}^{(k)} \times \mathbf{s}^{(k)}$
form an orthonormal basis for slip plane $\Pi^{(k)}$. 
Since  $\nabla \gamma^{{(k)}} \times \mathbf{n}^{{(k)}}$ is orthogonal to $\mathbf{n}^{(k)}$, it can be expanded in terms of $\mathbf{s}^{(k)}$ and $\mathbf{l}^{(k)}$,
\begin{equation}\label{decomposition_Burgers_tensor}
    \nabla \gamma^{(k)} \times \mathbf{n}^{(k)} = \bigl[\mathbf{l}^{(k)} \cdot (\nabla \gamma^{(k)} \times \mathbf{n}^{(k)})\bigr]\mathbf{l}^{(k)}+\bigl[\mathbf{s}^{(k)} \cdot (\nabla \gamma^{(k)} \times \mathbf{n}^{(k)})\bigr]\mathbf{s}^{(k)}\nonumber\\
= (-\mathbf{s}^{(k)} \cdot \nabla \gamma^{(k)})\mathbf{l}^{(k)} + (\mathbf{l}^{(k)} \cdot \nabla \gamma^{(k)})\mathbf{s}^{(k)}.
\end{equation}
Hence we can rewrite the Burgers tensor as
\begin{equation}\label{bugers_tensor_k}
\mathbf{G} = \sum_{k} 
\bigl[ 
(-\mathbf{s}^{(k)} \cdot \nabla \gamma^{(k)})\mathbf{l}^{(k)} \otimes \mathbf{s}^{(k)} +(\mathbf{l}^{(k)} \cdot \nabla \gamma^{(k)})\mathbf{s}^{(k)} \otimes \mathbf{s}^{(k)} 
\bigr]
= \sum_{k} \mathbf{G}^{(k)},
\end{equation}
where $\mathbf{G}^{(k)} \equiv (-\mathbf{s}^{(k)} \cdot \nabla \gamma^{(k)})\mathbf{l}^{(k)} \otimes \mathbf{s}^{(k)}+ (\mathbf{l}^{(k)} \cdot \nabla \gamma^{(k)})\mathbf{s}^{(k)} \otimes \mathbf{s}^{(k)}$ is the Burgers tensor for slip system $k$. $(\mathbf{G}^{(k)})^{\text{T}}\mathbf{e}$ is the contribution of slip system $k$ to the Burgers vector (per unit area) with unit normal $\mathbf{e}$.

When a dislocation contacts the interface, it aligns with the interface-slip plane intersection $\boldsymbol{\xi}_{\text{I}}^{(k)}$.
The total dislocation density of slip system $k$ near the interface can be expressed as
\begin{equation}\label{dd_cal1}(\mathbf{G}^{(k)})^{\text{T}}\boldsymbol{\xi}_{\text{I}}^{(k)}=\rho^{(k)}\mathbf{b}^{(k)}=\rho^{(k)}b^{(k)}\mathbf{s}^{(k)},
\end{equation}
where the  Burgers vector and slip direction $\mathbf{s}^{(k)}$ have the same directions. 
Substituting Eq.~\eqref{bugers_tensor_k} into Eq.~\eqref{dd_cal1}, the left hand side of Eq.~\eqref{dd_cal1} can be written as
\begin{equation}\label{dd_cal2}
(\mathbf{G}^{(k)})^{\text{T}}\boldsymbol{\xi}_{\text{I}}^{(k)}
=\bigl[\nabla\gamma^{(k)}\cdot(\xi^{(k)}_{s}\mathbf{l}^{(k)}-\xi^{(k)}_{l}\mathbf{s}^{(k)})\bigr]\mathbf{s}^{(k)},
\end{equation}
where we decompose the unit line direction vector as $\boldsymbol{\xi}_{\text{I}}^{(k)}=\xi^{(k)}_{s}\mathbf{s}^{(k)}+\xi^{(k)}_{l}\mathbf{l}^{(k)}$. 
Comparing Eq.~\eqref{dd_cal1} and Eq.~\eqref{dd_cal2}, we find the dislocation density of $k^{\text{th}}$ slip system near the interface as
\begin{equation}\label{dd_cal3}   \rho^{{(k)}}=\frac{\nabla\gamma^{(k)}}{b^{(k)}}\cdot(\xi^{(k)}_{s}\mathbf{l}^{(k)}-\xi^{(k)}_{l}\mathbf{s}^{(k)}).
\end{equation}
We  relate the dislocation flux to  crystal plasticity  variables by taking the time derivative of Eq.~\eqref{dd_cal3} 
\begin{equation}\label{df_1}
    \dot{\rho}^{(k)}=\frac{1}{b^{(k)}}(\xi^{(k)}_s \mathbf{l}^{(k)} \cdot \nabla \dot{\gamma}^{(k)}-\xi^{(k)}_{l} \mathbf{s}^{(k)} \cdot \nabla \dot{\gamma}^{(k)})
    =-\frac{1}{b^{(k)}} \text{Div}(\xi^{(k)}_l\dot{\gamma}^{(k)}\mathbf{s}^{(k)}-\xi^{(k)}_s\dot{\gamma}^{(k)}\mathbf{l}^{(k)}),
\end{equation}
where the dislocation flux is $\mathbf{J}^{(k)} \equiv \dot{\gamma}^{(k)}/b^{(k)}(\xi^{(k)}_l \mathbf{s}^{(k)}-\xi^{(k)}_s \mathbf{l}^{(k)})$. 
The magnitude and direction of the flux $\mathbf{J}^{(k)}$ are $\dot{\gamma}^{(k)}/b^{(k)}$ and  $\boldsymbol{\xi}_{\text{I}}^{(k)}=\xi^{(k)}_{s}\mathbf{s}^{(k)}+\xi^{(k)}_{l}\mathbf{l}^{(k)}$ (i.e., perpendicular to the dislocation line ). 
Then the interface BC Eq.~\eqref{dissipation4} can be expressed as
\begin{equation}\label{dissipation_general}
\left(\begin{array}{c}
\dot{\gamma}^{(1)} \\ 
\dot{\gamma}^{(2)} \\ 
\dot{\gamma}^{(3)} \\
\dot{\gamma}^{(4)}
\end{array}\right)
=
\left(\begin{array}{cccc}
L_{11} & L_{12} & L_{13} & L_{14}\\
       & L_{22} & L_{23} & L_{24}\\
            &   & L_{33} & L_{34} \\
\text{sym.}  &  &    & L_{44}
\end{array}\right)
\left(\begin{array}{c}
\tau^{(1)}\rho^{(1)} b^{(1)} \\ 
\tau^{(2)}\rho^{(2)} b^{(2)} \\ 
\tau^{(3)}\rho^{(3)} b^{(3)} \\
\tau^{(4)}\rho^{(4)} b^{(4)}
\end{array}\right),
\end{equation}
where the dislocation density on each slip system is obtained by Eq.~\eqref{dd_cal3} from which the plastic strain rate on each slip system is determined. 
The interface BC implies that the slip systems are more likely to accumulate plastic deformation when there is a large plastic strain gradient in the corresponding slip system near the interface.
The plastic strain rate on slip system $k$ is $\dot{\gamma}^{(k)}=J^{(k)}b^{(k)}=\rho^{(k)}v^{(k)}b^{(k)}$.  
In accordance with the conservation of  Burgers vector condition Eq.~\eqref{Burgersreaction}, the total net plastic strain rate at each reaction point must  be zero. 
This implies that our interface BC implicitly incorporates material compatibility at the interface during  dislocation reactions.

\subsection*{Numerical simulations}
We incorporate  crystal plasticity as follows.
The slip rate on slip system $\alpha$, $\dot{\gamma}^\alpha$, is formulated as a function of the resolved shear stress $\tau^\alpha$ and the critical shear stress:
    $\dot{\gamma}^\alpha=f(\tau^\alpha,\tau^{\alpha}_{\text{c}})$
and the evolution of the critical shear stress is dependent on the total shear $\gamma$ and shear rate $\dot{\gamma}^\alpha$:
    $\tau^\alpha_{\text{c}}=g(\gamma,\dot{\gamma})$.
We employ the rate-dependent kinetic law for FCC metal slip systems proposed by Hutchinson \citep{hutchinson1976bounds}: 
\begin{equation*}
    \dot{\gamma}^\alpha=\dot{\gamma}_0 \left| \frac{\tau^\alpha}{\tau^\alpha_{\text{c}}}  \right|^{\frac{1}{m}} \text{sgn}(\tau^\alpha),
\end{equation*}
where $\dot{\gamma}_0$ and $m$ are material parameters that determine the reference shear rate and  rate sensitivity. 
The strain hardening is characterized by the evolution of the strength through the incremental relation:
\begin{equation*}\label{hardening}
    \dot{\tau}^{\alpha}_{\text{c}}=\sum_{\beta=1}^{n} h_{\alpha \beta}\dot{\gamma}^{\beta},
\end{equation*}
where $h_{\alpha \beta}$ are the slip hardening moduli and the sum is over $n$ activated slip systems.
Here $h_{\alpha \alpha}$ and $h_{\alpha \beta}$ are the self- and latent-hardening moduli.
A simple form for the slip hardening moduli is~\citep{peirce1982analysis}
\begin{equation}\label{Peirce}
    h_{\alpha \beta}=q_{\alpha \beta} h_0 \text{sech}^2 \left|\frac{h_0 \gamma}{\tau_{\text{s}}-\tau_0} \right|, 
\end{equation}
where $h_0$ and $\tau_0$ are the initial hardening modulus and yield stress, $\tau_\text{s}$ is the stage I threshold stress and $\gamma$ is the Taylor cumulative shear strain on all slip system.
$q_{\alpha \beta}$ is a measure of latent hardening; it is commonly set to 1.0 for coplanar slip systems $\alpha$ and $\beta$, and between 0 and 1 for other slip systems.
Other types of models for slip hardening by replacing  Eq.~\eqref{Peirce} with other expressions  to describe, for example,  three stage hardening of crystalline materials~\citep{bassani1991latent,zarka1975constitutive}.
\begin{table}[htb]
 \centering
 \caption{CPFEM parameters for copper}
\renewcommand{\arraystretch}{1.3} 
\begin{tabular}{m{6cm} m{2cm}} 
 \hline
 Parameters & Values \\ 
 \hline
 Rate sensitivity exponent, $m$ & 0.1 \\ 
 Reference strain rate, $\dot{\gamma}_0$ & 0.001 $\text{s}^{-1}$ \\
 Initial hardening modulus, $h_0$ & 541.5 MPa \\
 Stage I threshold stress, $\tau_{\text{s}}$ & 109.5 MPa \\
 Initial yield stress, $\tau_0$  & 60.8 MPa \\ 
 Latent hardening moduli ratio, $q_{\alpha \beta}$ & 1.0 \\ 
 \hline
\end{tabular}
\label{table:1}
\end{table}

\begin{figure*}[htp]
\centering
\includegraphics[width=0.85\linewidth]{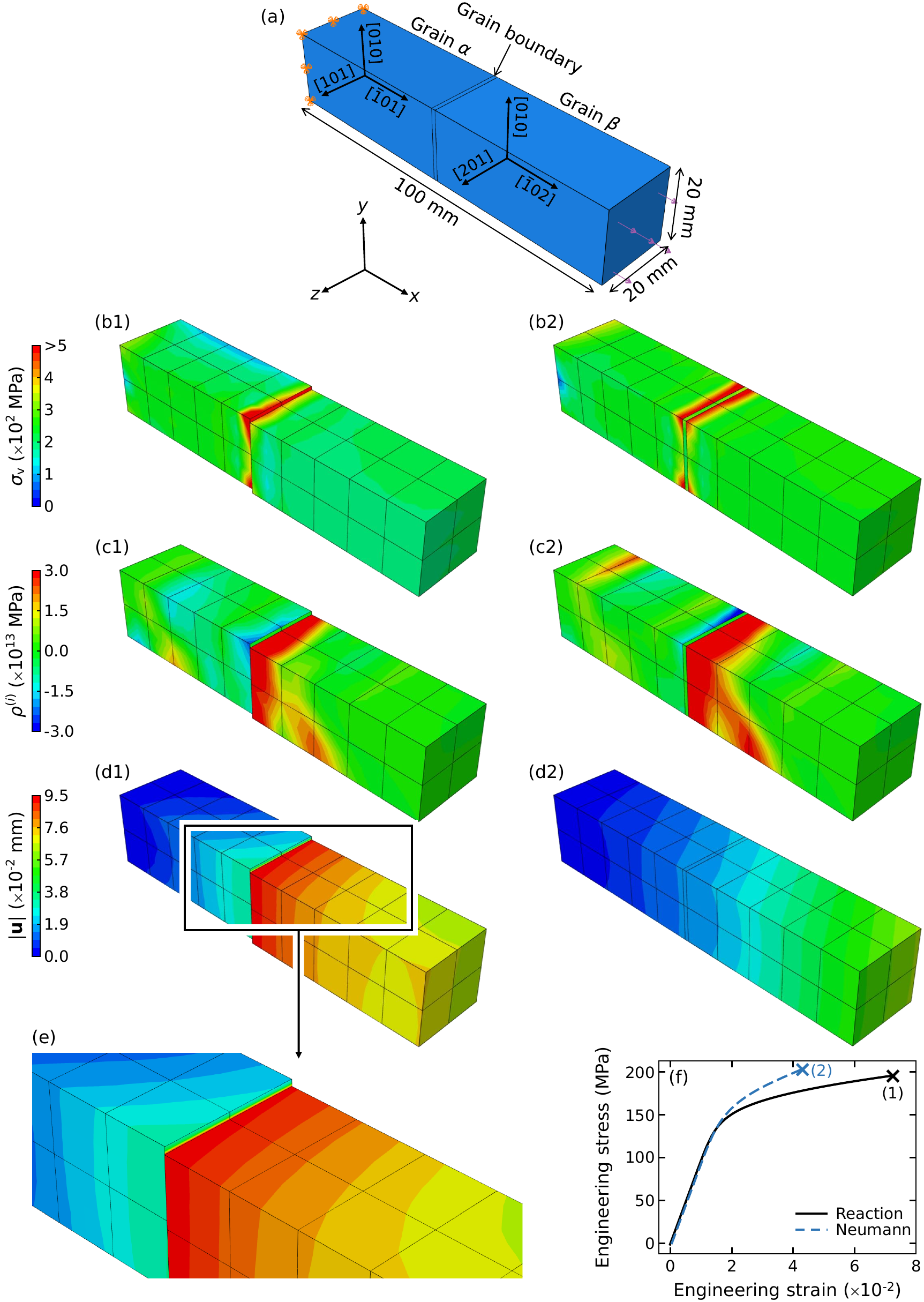}\hspace{-1.78em}%
\caption{\label{fig:cp_neumann}
(a) Schematic of finite element model;
(b) The contours of Von Mises stress, 
(c) dislocation density for one of slip systems and 
(d) displacement are plotted, the left and right-hand sides correspond to the Robin and Neumann BC respectively;
(e) The inset shows the sliding between two grains near the interface under Robin condition;
(f) Comparison of engineering stress-strain curve under Robin and Neumann BC condition.
}
\end{figure*}
This crystal plasticity model was incorporated into an Abaqus user material subroutine (UMAT) and applied to simulate the deformation of a $100$~mm long bar of square cross section ($20~\text{mm}  \times 20$~mm) of FCC copper with a set of activated $\{1 1 1\} \langle 1 1 0\rangle$ slip systems (see  Table \ref{table:1}).
The bar is a bicrystal (Grains $\alpha$ and $\beta$); see Fig.~\ref{fig:cp_neumann}(a). 
The GB is modeled as a thin region with a single $\{1 0 0\} \langle 0 0 1\rangle$ slip system.
The slip rate of the GB elements is calculated through the interface BC: $\dot{\gamma}^\txI=J^\txI b^\txI=\rho^\txI v^\txI b^\txI$.
The copper bicrystal bar is subject to uniaxial tensile stress of 200~MPa.
The displacements of all nodes at the left end plane of the bar are constrained as shown in Fig.~\ref{fig:cp_neumann}(a).
In this example, we assume that all dislocation reactions have the same  rates $\kappa_{(n)}$ in Eq.~\eqref{rbc10}.

Figure~\ref{fig:cp_neumann} shows the von Mises stress, dislocation density for one of the slip systems, and displacements. 
The left and right columns correspond to the  Reaction (Robin)  ($\kappa_{(n)} \neq 0$) and Impenetrable interface ($\kappa_{(n)} = 0$)  BC cases. 
As depicted in Fig.~\ref{fig:cp_neumann}(b), the von Mises stress contour reveals stress concentrations near the grain boundary (GB). 
Figure~\ref{fig:cp_neumann}(c) demonstrates the accumulation of dislocations with opposite signs near the GB for one of the activated slip systems.
Figure~\ref{fig:cp_neumann}(d1) shows that the displacement field is discontinuous across the Reaction BC GB due to GB sliding. 
Comparing the Reaction and Impenetrable cases, we see that dislocation reactions near the interface relax  the high stress at the interface (Impenetrable BC case) by allowing some dislocation transmission. 
This is responsible for the smaller von Mises stress concentrations near the Reaction BC GB than in the Impenetrable BC  case. 
It is especially interesting to note that the Reaction BC results in GB sliding, even when the bar is axially loaded, as shown in Fig.~\ref{fig:cp_neumann}(e).
Because the Reaction BC allows for some slip transfer across the GB than when the GB is impenetrable, the Reaction BC bicrystal shows a lower yield strength and is less hardening than when it is impenetrable, as illustrated in Fig.~\ref{fig:cp_neumann}(f). 

\section{Application to Discrete Dislocation Dynamics}\label{DDD}
When an individual  dislocation approaches an interface, it may be blocked, transmitted,  or reflected onto another slip system.
In most  DDD simulations,  dislocations were able to move on new slip systems according to empirical, geometric, or thermodynamic rules -- often drawn from  experimental or molecular simulation observations~\citep{lu2019grain,aragon2022dislocation}. 
Here, we explicitly account for dislocation reactions at the interface based on our Reactive BC interface model and all possible slip systems.
\begin{figure*}[hbt]
\centering
\includegraphics[width=1\linewidth]{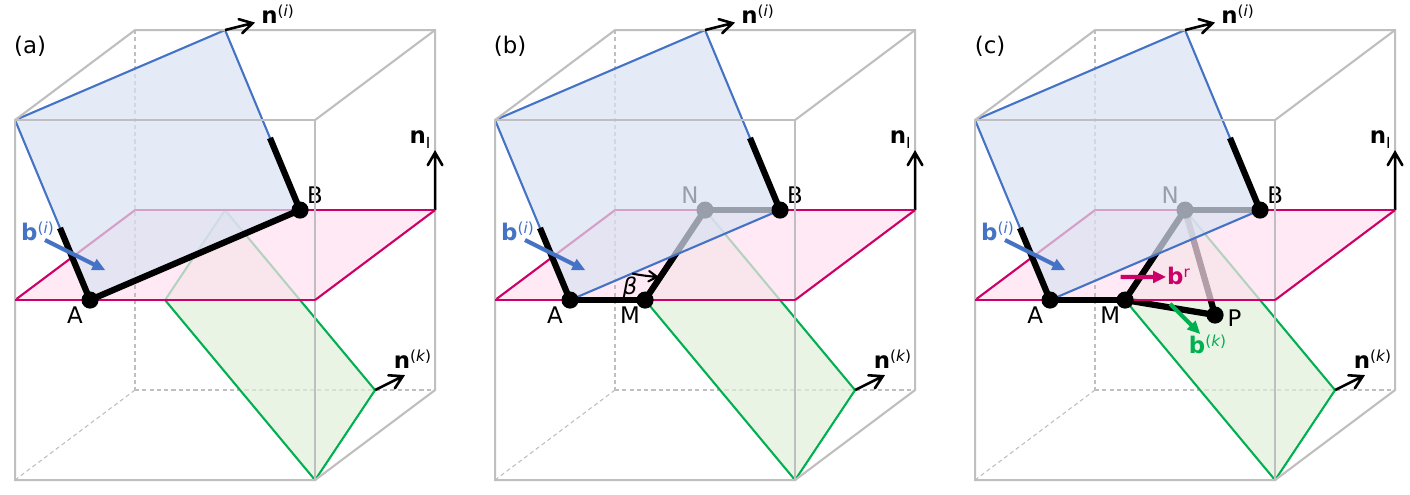}\hspace{-1.78em}%
\caption{\label{fig:ddd_schematic}
Schematic of transmissions of dislocation in DDD. (a) Single dislocation line from $i^{\text{th}}$ slip plane is approaching interface; (b) the dislocation line rotate by $\beta$ angle; (c) dislocation nucleation of $k^{\text{th}}$ slip system.
}
\end{figure*}

In DDD simulations, each dislocation is discrete in the reaction process. 
Assuming that there is single dislocation line from slip system $i$ approaching the interface as shown in Fig.~\ref{fig:ddd_schematic}(a) .
The dislocation density of incoming slip system is $\rho^{(i)}=1/A_{\text{e}}$, where $A_{\text{e}}$ is the unit area near the interaction. 
There are many potential outgoing slip system from which  to choose, corresponding to the many possible reactions amongst the different slip systems. 
For an arbitrary reaction, denoted by  ``$n$'', we can substitute dislocation density $\rho^{(i)}$ into the interface BC Eq.~\eqref{dissipation4} and obtain the dislocation fluxes of reaction $n$,  $\mathbf{J}_{n}=(J^{(i)}_{(n)},J^{(j)}_{(n)},J^{(k)}_{(n)},J^{(l)}_{(n)})$.
The fluxes of all  ($C_{N-1}^{3} \equiv m$) possible reactions with slip system $i$  involved can be calculated following the same procedure  
\begin{equation}
     \left\{
     \begin{array}{l}
          (J^{(i)}_{(1)},J^{(1)}_{(1)},J^{(2)}_{(1)},J^{(3)}_{(1)}) \\
          (J^{(i)}_{(2)},J^{(1)}_{(2)},J^{(2)}_{(2)},J^{(4)}_{(2)}) \\
          ... \\
          (J^{(i)}_{(m)},J^{(j)}_{(m)},J^{(k)}_{(m)},J^{(l)}_{(m)})
     \end{array}\right..
 \end{equation}
In an energetically favorable structure, the nucleated dislocation should form such that it dissipates the store elastic energy at the fastest rate (this is equivalent to the maximum rate of entropy production)~\citep{cho2020dislocation}.
We then determine the outgoing slip system following two steps:

\begin{itemize}[align=left, leftindent=0em, left margin=0em, itemindent=!, nosep, noitemsep]
\item [{1.}]
Calculate the reaction rates of all possible reactions.
\item[{2.}]
Choose the reaction with the maximum power dissipation amongst all possible reactions.
Since the interface Reaction BC is based upon  linear response theory, it corresponds to evolution in the direction that maximizes the rate of entropy production.
The magnitude of the local entropy production rate can be expressed as the product of the driving force and dislocation flux for each slip system~\citep{yu2023mesoscale}.
The local entropy production rate for reaction ``$n$''  is the sum of the entropy production rate over all four slip systems involving in this reaction ($\dot{s}_{(n)}=\sum_{k=1}^{4}|f^{(k)}J^{(k)}_{(n)}|$).
The dominant reaction is that ``$n$'' with the largest entropy production rate (amongst all reactions $\dot{s}_{(n)}$) and the outgoing system $k$ with the largest entropy production rate $\dot{s}^{(k)}_{(n)}$ (amongst the three ``outgoing'' slip systems).
\end{itemize}
Once the outgoing slip system is determined, the dislocation line MN on slip system $i$ rotates by $\beta$  to align with slip plane $k$ as Fig.~\ref{fig:ddd_schematic}(b) shows.
The residual Burgers vector left at the interface is
    $\mathbf{b}^{r}=\mathbf{b}^{(i)}-\mathbf{b}^{(k)}$.
The velocity of the emitted dislocation on the  outgoing slip plane  $k$ is
    $v^{(k)}=J^{(k)}_{(n)}/\rho^{(k)}$,
where dislocation density on slip system $k$ is $\rho^{(k)}=1/A_{\text{e}}$, the constant  unit area $A_{\text{e}}$ cancels.
The new segment will bow out along slip plane $k$   as shown in Fig.~\ref{fig:ddd_schematic}(c) . 
The new node will be created and the distant between line MN and new node P is
    $d^{(k)}=v^{(k)} \rmd t$,
where $\rmd t$ is the time step of the iteration. 

\section{Discussion}
In the  simulations discussed above, we made the simplifying assumption that the interface is flat, while  interfaces are  often curved. 
To address this, the distribution of disconnections (line defects with Burgers vector and step character) in the interface should be considered~\citep{han2022disconnection}. 
Specifically, both the shape of any interface and the relative displacements between the phases/grains can be rigorously described in terms of  the distribution of disconnection steps and Burgers vectors, respectively. 
Since the step heights are not involved in interfacial Burgers vector reactions, the interface boundary condition (BC)  can still be applied to a curved interface by disregarding the disconnection steps at each point.
However, some modifications are necessary when applying the  interface BC to a curved interface since its fundamental character varies from point-to-point. 
First, the geometry term $(\mathbf{c} \otimes \mathbf{c})$ in Eq.~\eqref{dissipation4} is related to $\{ \mathbf{s}^{\txI}, \mathbf{n}^{\txI} \}$, and its tangential and normal directions will change along the curved interface plane. 
Second, the driving force along the interface, $\tau^\txI$, and the reaction constant, $\kappa$, may vary with position along the curved interface.

The variation of reaction constants is necessary for different dislocation reactions. 
In our theory, the reaction constant reflects the energy  dissipation rate for the dislocation reactions near the interface; this is controlled by  microscopic mechanism (the atomic structure of the interface).
This is similar to the Gurtin  theory for GBs~\citep{gurtin2008theory} which is based on a reduced dissipation inequality of the form 
\begin{equation*}\label{gurtin_2st}
    \mathbb{D}=\mathbb{K} : \dot{\mathbb{G}} \geq 0,
\end{equation*}
where $\mathbb{K}$ is a tensor internal force defined over the region and $\dot{\mathbb{G}}$ measures the rate of defect accumulation at the interface.
In order to satisfy the above inequality, the $\mathbb{K}$ should take the form, $\mathbb{K}=\mathbb{F}:\dot{\mathbb{G}}$, where $\mathbb{F}$ is a fourth-order positive definite tensor.
Gurtin proposed a simple dissipative constitutive relation in which  $\mathbb{F}$ is proportional to the identity tensor, i.e, $\mathbb{F}=F\mathbb{I}$, 
\begin{equation*}\label{Gurtin_dis}
  \mathbb{K} \equiv F |\dot{\mathbb{G}}|^m \frac{\dot{\mathbb{G}}}{|\dot{\mathbb{G}}|},
\end{equation*}
where $m>0$ is a rate-sensitivity modulus and $F>0$ is a hardening-softening parameter that reflects the energy dissipation rate, similar to $\kappa_{(n)}$ in our theory.
This parameter can only  be determined based upon a microscopic mechanism (and atomistic simulation).
The climbing-image nudged elastic band (CINEB) method~\citep{zhu2007interfacial} may be employed to calculate the energy barrier $Q$ for specific dislocation reactions. 
Then, by employing the equation $\kappa=\kappa_0 \exp(-Q/k_{\text{B}}T)$, where $Q$ and $k_{\text{B}}T$ are the energy barrier and  thermal energy, we can, in principle,  determine the reaction constant $\kappa$. 
However, considering the numerous reactions (and reaction constants) occurring among all slip systems, this is not a practical solution. 

In polycrystalline materials, plastic deformation is most frequently observed as a macroscopic average behavior across a large  (compared to the atomic scale) spatial, polycrystalline region. 
Therefore, for simplicity, we can assume that all reaction constants may be ``lumped'' together, resulting in a single constant, as done in this paper. 
However, determining the specific reaction constant for simulations poses a question. 
One possible approach is to treat it as a parameter and  calibrate it by fitting to experimental stress-strain data. 
Another approach for determining the reaction constant is to measure the activation volume $\Omega \equiv -(\partial Q / \partial \sigma)_T$. 
At constant temperature $T$ and strain rate $\dot{\epsilon}$, the activation energy $Q$ may be expressed as~\citep{zhu2008temperature}:
\begin{equation*}\label{activation_energy}
    \frac{Q}{k_{\text{B}}T}=\ln \frac{k_\text{B} TN\nu_0}{E\dot{\epsilon}\Omega},
\end{equation*}
 where $N$ and $\nu_0$ are known parameters.  
For a tensile test, the activation volume is $\Omega=k_{\text{B}}T\partial \ln\dot{\epsilon}/\partial \sigma$, where $\dot{\epsilon}$ and $\sigma$ represent the strain rate and stress.

The aforementioned approaches represent two extreme cases. 
On the one hand, it is impractical to calculate each reaction constant individually due to the numerous reactions occurring in polycrystalline materials and the computational cost of determining each. 
On the other hand, assuming identical reaction constants for all GBs oversimplifies the analysis. 
In order to obtain more accurate reaction constants without excessive computational resources, other improvements may be implemented.
One possible approach is to utilize the nudged elastic band (NEB) method to calculate the activation energy for various types of GBs. 
Initially, the calculated GBs can be categorized into three distinct types: high-angle, low-angle, and coherent-twin GBs, as their activation energies differ significantly. 
These can be grouped, thereby reducing the number of  activation energies determinations required.
Subsequently, these reaction constants may be assigned to different GBs based on their respective proportions in the polycrystalline material.

\section{Conclusion}
In this paper, we proposed and demonstrated how our rigorous, conservation of Burgers vector-based, interface BC can be combined with different simulation methods to examine plastic deformation in polycrystalline and/or multiphase materials:
\begin{itemize}
\item[(i)] 
In continuum dislocation dynamics, the dislocation flux at the interface may be updated by the interface boundary condition.
The numerical examples show that the reflection and transmission of dislocations at the interface can be captured by our interface boundary condition.
\item[(ii)] 
In the crystal plasticity finite element method, the plastic strain rate of one slip system near the interface will be coupled with the gradient of plastic strain of all activated slip systems. 
A bicrystal model simulation results show that the pile-ups of dislocation and stress concentration near the interface are weakened by the permeability of the interface to dislocations.
Interface sliding is a natural consequence of interfacial dislocation reactions and is captured by this approach.
\item[(iii)]
For discrete dislocation dynamics, possible dislocation reactions at the interface are diverse. 
The outgoing slip system can be determined based upon a maximum rate of entropy production criterion, which can be expressed as the product of the dislocation  flux and driving force on each slip system participating in the interfacial reactions.
\end{itemize}

The successful implementation of our mesoscale interface boundary condition within the continuum dislocation dynamics and crystal plasticity finite element method has been validated through simulations. 
This makes it possible to apply the interface boundary condition to a wide-range of numerical plasticity approaches following similar procedures.

\section*{Acknowledgements}
JY, DJS and JH were supported by the National Key R\&D Program of China (2021YFA1200202). 
JY, AHWN and DJS also gratefully acknowledge support of the Hong Kong Research Grants Council Collaborative Research Fund C1005-19G. 
JH acknowledges support from the Early Career Scheme (ECS) Grant of the Hong Kong Research Grants Council CityU21213921. 
AHWN also acknowledges support from the Shenzhen Fund 2021 Basic Research General Programme JCY20210324115400002 and the Guangdong Province Basic and Applied Research Key Project 2020B0301030001.

\bibliographystyle{elsarticle-harv}
\bibliography{mybib}
\end{document}